\documentclass[12pt]{article}

\def\fnote#1{\footnote}

\def\a,{a\kern-.28em\rlap{\lower.80ex\hbox{${}_`$}}\kern.28em\relax}
\def\e,{e\kern-.25em\rlap{\lower.80ex\hbox{${}_`$}}\kern.25em\relax}
\def\A,{A\kern-.26em\rlap{\lower.95ex\hbox{${}_`$}}\kern.26em\relax}
\def\E,{E\kern-.28em\rlap{\lower.95ex\hbox{${}_`$}}\kern.28em\relax}
\def\x,{\.{z}}
\def\X,{\.{Z}}

 \usepackage[dvips]{graphicx}

\begin{document}

\title{Structure formation in the Lema\^{\i}tre-Tolman model}

\author{
Andrzej
Krasi\'nski\thanks{This research was supported by the Polish Research
Committee grant no 2 P03B 060 17 and by a grant from the South African
National Research Foundation}\\
N. Copernicus Astronomical Centre, Polish Academy of Sciences, \\
Bartycka 18, 00 716 Warszawa, Poland, email: akr@camk.edu.pl
\and
Charles Hellaby$^*$ \\
Department of Mathematics and Applied Mathematics, \\ University of
Cape Town
\\ Rondebosch 7701, South Africa, email: cwh@maths.uct.ac.za
}

\date {}

\maketitle

 \begin{abstract}
 Structure formation within the Lema\^{\i}tre-Tolman model is investigated
in a general manner.  We seek models such that the initial density
perturbation within a homogeneous background has a smaller mass than the
structure into which it will develop, and the perturbation then accretes
more mass during evolution.  This is a generalisation of the approach
taken by Bonnor in 1956.  It is proved that any two spherically symmetric
density profiles specified on any two constant time slices can be joined
by a Lema\^{\i}tre-Tolman evolution, and exact implicit formulae for the
arbitrary functions that determine the resulting L-T model are obtained.
Examples of the process are investigated numerically.
 \end{abstract}

 \begin{center}
 {\tt gr-qc/0106096} \\[4mm]
 {\it Phys. Rev. D, Submitted 29/6/01, Accepted 30/8/01} \\[4mm]
 PACS: 
 \parbox[t]{80mm}{
 98.80.-k Cosmology, \\
 98.62.Ai Formation of galaxies, \\
 98.65.-r Large scale structure of the universe 
 }
 \end{center}

 \section{Introduction}

 \setcounter{equation}{0}

 Though the Lema\^{\i}tre-Tolman (L-T) model has been studied extensively
(see \cite{Kras1997}), the question of whether galaxies and other modern
cosmic structures can grow from small initial perturbations, using this
exact inhomogenous cosmological model, has not been clearly answered.

 Long ago Bonnor \cite{Bonn1956} considered a version of the general
problem, in which the perturbation consisted of an interior region
matched to a Friedmann-Lema\^{\i}tre-Robertson-Walker (Friedmann)
exterior, so the initial fluctuation of density included all the
dust particles that would enter the future galaxy, and so the
outer edge of the perturbation had to be comoving with the
(spatially homogeneous) background flow ever after. Because of
this, and given the present age of the Universe, the initial
fluctuation had to have an amplitude many times larger than a
statistical fluctuation could have.  However current matter models, that 
allow perturbations to grow before recombination, have successfully 
predicted temperature perturbations in the CMB of order $10^{-5}$.

 The very existence of inhomogeneous cosmological models (i.e. spatially
inhomogeneous solutions of Einstein's equations with expanding matter),
such as the L-T \cite{Lema1933, Tolm1934} or Szekeres \cite{Szek1975,
GoWa1982} models, shows that non-Friedmannian distributions of density and
velocity would have been coded in the Big Bang and need not be
``explained" as statistical fluctuations that appeared within a
homogeneous background during evolution. Moreover, since the L-T
collection of models is labelled by two arbitrary functions of mass, that
reduce to specific forms in the Friedmann limit, it follows that the dust
Friedmann models are a subset of measure zero within the L-T set.
Consequently, the Friedmann models are very improbable statistically and,
assuming that our physical Universe is homogeneous indeed, one needs to
explain how homogeneity might have come about out of inhomogeneous initial
data, not the other way round. This is what inflation is believed to have
done. However, in this paper we shall accept a high degree of homogeneity
at decoupling, and we will determine how fast condensations can grow, once
they appear in a homogeneous background.

 In this paper we shall relax Bonnor's assumptions, and consider the
general case.  In particular, we envisage a scenario in which the mass of
the initial fluctuation is much smaller than the mass of the condensation
into which it will develop, and that it captures more mass during its
evolution.\footnote{Such a modification of Bonnor's method was suggested
by S. Ba\x,a\'nski during one of the seminars by A. K. in Warsaw.} The
outer edge of the growing condensation will thus not be comoving with the
background flow. The calculations, although based on exact formulae, will
have to be carried out numerically.

 In the following we set $\Lambda = 0$, as its effect is felt primarily at
late times over the long range, and so will not strongly affect structure
formation.  Also, current interpretations of the CMB and Supernova data
that estimate a non-zero $\Omega_\Lambda$, should be regarded as
provisional, since several reasonable alternatives have been put forward.

 \section{Basic properties of the Lema\^{\i}tre-Tolman model.}

 \setcounter{equation}{0}

The Lema\^{\i}tre-Tolman (L-T) model \cite{Lema1933, Tolm1934} is
a spherically symmetric nonstatic solution of the Einstein
equations with a dust source. Its metric is:

\begin{equation}
{\rm d}s^2 = {\rm d}t^2 - \frac {{R,_r}^2}{1 + 2E(r)}{\rm d}r^2 -
R^2(t,r)({\rm d}\vartheta^2 + \sin^2\vartheta{\rm d}\varphi^2),
\end{equation}

\noindent where $E(r)$ is an arbitrary function (arising as an
integration constant from the Einstein equations), $R,_r$ is the
derivative of the function $R(t,r)$ by $r$, and $R$ obeys the
equation

\begin{equation}   \label{Rtsq}
{R,_t}^2 = 2E + 2M/R + \frac 13 \Lambda R^2,
\end{equation}

\noindent where $\Lambda$ is the cosmological constant. Eq.~(\ref{Rtsq}) is
a first integral of one of the Einstein equations, and $M(r)$ is another
arbitrary function that arises as integration constant. The matter-density
is:

\begin{equation}   \label{rhoLT}
\kappa \rho = \frac {2M,_r}{R^2R,_r}, \qquad {\rm where}\ \kappa =
\frac {8\pi G} {c^4}.
\end{equation}

In the following, we will assume $\Lambda = 0$. Then eq.~(\ref{Rtsq}) can
be solved explicitly. The solutions are:

When $E < 0$:

\begin{equation}   \label{EllEv}
R(t,r) = - \frac{M}{2E}(1 - \cos\eta),$$ $$\eta - \sin\eta = \frac
{(-2E)^{3/2}}{M} (t - t_B(r)).
\end{equation}

\noindent where $\eta$ is a parameter; when $E = 0$:

\begin{equation}   \label{ParEv}
R(t,r) = \left[ \frac{9}{2} M (t - t_B(r))^2\right]^{1/3},
\end{equation}

\noindent and when $E > 0$:

\begin{equation}   \label{HypEv}
R(t,r) = \frac{M}{2E}(\cosh\eta - 1),$$ $$\sinh\eta - \eta = \frac
{(2E)^{(3/2)}}{M} (t - t_B(r)),
\end{equation}

\noindent  where $t_B(r)$ is one more arbitrary function (the bang
time). Note that all the formulae given so far are covariant under
arbitrary coordinate transformations $r = g(r')$, and so $r$ can
be chosen at will. This means one of the three functions $E(r)$,
$M(r)$ and $t_B(r)$ can be fixed at our convenience by the
appropriate choice of $g$.

In a general L-T model, $E$ may change sign, having both re-collapsing
and ever-expanding regions.  Also the space $t =$ const with $E(r) < 0$
everywhere is not necessarily closed (and the one with $E(r) > 0$ is not
necessarily infinite), see Refs. \cite{HeLa1985, Bonn1985}.

The Friedmann models are contained in the Lema\^{\i}tre-Tolman class as
the limit:

\begin{equation}
t_B = {\rm const}, \qquad |E|^{3/2}/M = {\rm const},
\end{equation}

\noindent and one of the standard radial coordinates for the Friedmann 
model results if the coordinates in (\ref{EllEv}) -- (\ref{HypEv}) are 
chosen so that: 

\begin{equation}
M = M_0 r^3,
\end{equation}

\noindent where $M_0$ is an arbitrary constant; so that $E =
E_0r^2,\ E_0 =$ const.

It will be convenient in most of what follows to use $M(r)$ as the radial 
coordinate (i.e. $r' = M(r)$) because in the structure formation context 
one does not expect any ``necks" or ``bellies" where $M,_r = 0$, and so 
$M(r)$ should be a strictly growing function in the whole region under 
consideration.  (See the papers by Barnes and, especially, by Hellaby 
\cite{Barn1970, Hell1987} for descriptions of necks.  Some properties of a 
neck appeared also in the paper by Novikov \cite{Novi1962}).  Then: 

\begin{equation}   \label{rhoRM}
\kappa \rho = 2/(R^2R,_M) \equiv 6/(R^3),_M.
\end{equation}

 We also, in searching for realistic models, prefer L-T models that are
free of shell crossings \cite{HeLa1985}.  This is not because shells of
matter cannot collide, but because the L-T co-moving description breaks
down there.

 We shall now apply this model to the problem of structure formation
within the exact relativity theory (i.e. without approximations).  We
believe this question has not been satisfactorily answered so far, and so
it deserves to be investigated more thoroughly, both as an important
consequence of L-T models, and with a view to possible cosmological
applications.  We seek accreting models in which a small initial
fluctuation at decoupling captures more mass during its evolution, thus
growing in extent as well as in density contrast.

 \section{The evolution as a mapping from an initial density to a final
density.}
 \label{TheoremSection}

 \setcounter{equation}{0}

The evolution of the L-T model is usually specified by defining
the initial conditions --- the distributions of the Big Bang time
$t_B(M)$ and of energy $E(M)$, or by specifying e.g. the density
$\rho(t_1, R)$ and velocity $R_{,t}(t_1, R)$ at an initial instant
$t = t_1$. It is, however, possible, to approach the problem in a
different way: to specify the density distributions at two
different instants, $t = t_1$ and $t = t_2$, calculate the
corresponding $E(M)$ and $t_B(M)$, and in this way obtain a
definite model. It is not immediately obvious whether all pairs of density 
distributions may be connected by an L-T evolution of a chosen type; nor 
whether one can ensure shell-crossings do not occur between $t_1$ \& 
$t_2$. However, such a mapping from an initial density to a given final 
density should exist in many cases, especially with a sensible choice of 
density profiles. 

In fact, it will be proven below that any initial value of density at a 
specific position ($r,M$ const) can be connected to any final value of 
density at the same position by one of the Lema\^{\i}tre-Tolman evolutions 
(either $E > 0$, or $E < 0$, or, in an exceptional case, $E = 0$). In the 
Friedmann limit, any two constant densities can be connected by one of the 
$k > 0$, $k < 0$ or $k = 0$ Friedmann evolutions. 

 For definiteness, it will be assumed in the following that the final 
instant $t_2$ is later than the initial instant $t_1$, i.e. $t_2 > t_1$, 
and that the final density $\rho(t_2, M)$ is smaller than the initial 
density $\rho(t_1, M)$ at the same $M$.  We thus assume that matter has 
expanded along every 
 world-line, but the proof can be easily adapted to the collapse 
situation. 

 \subsection{Hyperbolic regions}

Let us consider the L-T model with $E > 0$.  Let the initial and final 
density distributions at $t = t_1$ and $t = t_2$ be given by: 

\begin{equation}
  \rho(t_1,M) = \rho_1(M), ~~~~~~~~~~~~
  \rho(t_2,M) = \rho_2(M)
\end{equation}

\noindent From (\ref{rhoLT}) we then have, for each of $t_1$ \& $t_2$:

\begin{equation}   \label{Rintegral}
 R^3(t_i, M) - R^3_{{\rm min}\,i} = \int_{M_{\rm min}}^M \frac{6}{\kappa 
\rho_i(M')} \, {\rm d}M' := {R_i}^3(M), ~~~~~~~~i = 1,2
\end{equation} 

\noindent and $R_2(M) > R_1(M)$ in consequence of $\rho(t_2,M) <
\rho(t_1,M)$.  In the following we will assume there is an origin where $M 
= 0$ and $R(t_i, 0) = 0$, so that $R_{{\rm min}\,i} = 0 = M_{\rm min}$ is 
valid%
 \footnote{
 For examples where this is not the case, see the papers by Einstein \& 
Straus \cite{EiSt1945,EiSt1946} with $M_{min} \neq 0$ \& $R_{min} = 0$, 
and by Bonnor \& Chamorro \cite{BoCh1990} where $M = 0$ from $R = 0$ to 
$R_{min}$.  Also ``necks" are the locus of a minimum in $M$ and $R$.
 }%
 .  Solving the two parts of (\ref{HypEv}) for $t(R,r)$ and writing it out 
for each of $(t_1, R_1)$ and $(t_2, R_2)$ leads to: 

\begin{equation}   \label{tBiHyp}
t_B = t_i - \frac {M} {(2E)^{3/2}} \left[\sqrt{(1 + 2ER_i/M)^2 -
1} - {\rm arcosh}(1 + 2ER_i/M)\right], ~~~~~~~~i = 1,2
\end{equation}

\noindent and then eliminating $t_B$ between the two versions of 
(\ref{tBiHyp}) we find: 

\begin{equation}   \label{t2t1Hyp}
\sqrt{(1 + 2ER_2/M)^2 - 1} - {\rm arcosh}(1 + 2ER_2/M) $$ $$ -
\sqrt{(1 + 2ER_1/M)^2 - 1} + {\rm arcosh}(1 + 2ER_1/M) =
[(2E)^{3/2}/M](t_2 - t_1),
\end{equation}

\noindent We shall prove that this equation has one and only one solution 
$E(M) > 0$ (in addition to the trivial solution $E = 0$) provided that 
$t_2$ and $t_1$ obey a certain inequality (see below). In fact, the 
inequality will exclude the $E \leq 0$ models. 

For ease of calculations, let us denote:

\begin{equation}   \label{psiH}
x := 2E/M^{2/3}, \qquad a_i = R_i/M^{1/3}, \ \ i = 1,2; $$ $$
\psi_H(x) := \sqrt{(1 + a_2x)^2 - 1} - {\rm arcosh}(1 + a_2x) -
\sqrt{(1 + a_1x)^2 - 1} + {\rm arcosh}(1 + a_1x) $$ $$- (t_2 -
t_1)x^{3/2} := \chi_H(x) - (t_2 - t_1)x^{3/2},
\end{equation}

\noindent Our problem is then equivalent to the following question: for
what values of the parameters $a_2 > a_1$ and $t_2 > t_1$, does
the equation $\psi_H(x) = 0$ have a solution $x \neq 0$?  Note that 
$\psi_H$ has a zero at $x = 0$ that the more correct but less convenient 
$x^{-3/2} \chi_H - (t_2 - t_1)$ does not have. 

Since $x \geq 0$ and $a_i > 0$ by definition, $\psi_H(x)$ and
$\chi_H(x)$ are well-defined for any $x \in [0, \infty)$. Note that

\begin{equation}   \label{LimxInfHyp}
\lim_{x \to \infty}\frac{{\rm arcosh}(1 + a_ix)} {\sqrt{(1 +
a_ix)^2 - 1}} = \lim_{x \to \infty}\frac{{\rm arcosh}(1 + a_ix)}
{(t_2 - t_1)x^{3/2}} = 0,
\end{equation}

\noindent and so in determining the sign of $\psi_H(x)$ as $x \to
\infty$, the arcosh-terms can be neglected. Note also

\begin{equation}
\lim_{x \to \infty}\frac{\chi_H(x)} {(t_2 - t_1)x^{3/2}} = 0.
\end{equation}

\noindent Hence, the last term in $\psi_H$ becomes dominant when $x
\to + \infty$, and so

\begin{equation}
\lim_{x \to +\infty}\psi_H(x) = - \infty.
\end{equation}

\noindent It is easy to see that $\psi_H(0) = 0$, but we wish to
know whether $\psi_H \to 0^+$ or $\psi_H \to 0^-$ as $x \to 0$. For
this purpose, note that

\begin{equation}   \label{dpsidxHyp}
\frac{d}{dx}\psi_H(x) = \psi_{H,x} = \sqrt{x} \left[\frac 
{{a_2}^{3/2}} {\sqrt{2 + a_2x}} -\frac {{a_1}^{3/2}} {\sqrt{2 + a_1x}} - 
\frac 32 (t_2 -t_1)\right],
\end{equation} 

\noindent from which it follows that

\begin{equation}
\psi_{H,x}(0) = 0, \qquad \lim_{x \to \infty} \psi_{H,x} = - \infty.
\end{equation}

\noindent It is also easy to see that the term in square brackets
in (\ref{dpsidxHyp}) is a strictly decreasing function for all $x \in [0,
\infty)$, and so it may be equal to zero in at most one point.
Since it goes to the negative value $- (3/2)(t_2 - t_1)$ when $x
\to \infty$, it will have a zero when it is positive at $x = 0$,
i.e. when

\begin{equation}   \label{HypCondit}
t_2 - t_1 < \frac {\sqrt{2}}3 \left({a_2}^{3/2} -
{a_1}^{3/2}\right).
\end{equation}

\noindent By comparison with (\ref{ParEv}), since $a_i = R_i/M^{1/3}$, the 
above is seen to be equivalent to the statement that between $t_1$ and 
$t_2$, $R(t, M)$ has increased by more than it would have increased in the 
$E = 0$ L-T model. This is a necessary condition for the existence of an 
$E > 0$ evolution connecting $R(t_1, M)$ to $R(t_2, M)$.

It is also a sufficient condition, as we now explain.  With 
(\ref{HypCondit}) fulfilled, $\lim_{x \to 0} \psi_{H,x} = 0^+$, i.e. 
$\psi_{H,x} > 0$ in a neighbourhood of $x = 0$, then it goes through zero 
exactly once, at some $x = x_m$, and becomes negative. This means that 
$\psi_H(x)$ itself is increasing from the value 0 at $x = 0$, to a maximum 
at $x = x_m$, and is then decreasing all the way to $x \to \infty$ where 
it becomes $- \infty$. Hence, at one and only one $x = x_0 > x_m$, 
$\psi_H(x_0) = 0$. This implies that eq.~(\ref{t2t1Hyp}) defines a function 
$E(M)$ in the whole range of $M$ in which (\ref{HypCondit}) is fulfilled. 
Examples of the functions $\psi_H(x)$ from (3.8) that obey or do not obey 
(\ref{HypCondit}) are shown in Fig. 1. 

 \begin{center}
 --------------------------- \\
 Fig. 1. goes here \\
 ---------------------------
 \end{center}

There remains a practical problem for the numerical calculation of
$E(M)$. Since the range of $x$ is infinite in (\ref{psiH}), an
initial value $x_A < \infty$ such that $\psi_H(x_A) < 0$ has to be
determined first. For this purpose, note that for large $x$ we
have $\sqrt{(1 + a_ix)^2 - 1} \approx 1 + a_ix$. Together with
(\ref{LimxInfHyp}), this implies that $\psi_H(x)$ is well approximated for
large $x$ by

\begin{equation}   \label{psiA}
\psi_{A}(x) := (a_2 - a_1)x - (t_2 - t_1)x^{3/2}.
\end{equation}

\noindent Indeed, it is easy to verify that for all $x > 0$

\begin{equation}
\psi_H(x) < \psi_{A}(x).
\end{equation}

\noindent [Writing $\psi_H(x) - \psi_{A}(x) \equiv \chi_H(x) - (a_2 -
a_1)x$, and $(\psi_{H,x} - \psi_{A,x}) = W(a_2) -W(a_1)$, where $W(a) = 
a^{3/2}\sqrt{x}/\sqrt{2 + ax} - a$, we note that $W(a)$ is a decreasing 
function of $a$ for every $x > 0$, while $\psi_H(0) = \psi_{A}(0) = 0$. 
Hence, $\psi_H(x) < \psi_{A}(x)$ for $x > 0$.] Therefore, if 
$\psi_{A}(x_A) = 0$, then $\psi_H(x_A) < 0$. 

The solution of $\psi_{A}(x) = 0$ is

\begin{equation}
x_A = \frac {(a_2 - a_1)^2} {(t_2 - t_1)^2},
\end{equation}

\noindent and so $x_A$ is a good initial value for the numerical
program that will find a solution of $\psi_H(x) = 0$ by bisecting
the segment $[0, x_A]$ and checking the sign of $\psi_H(x_A/2)$.

Also, for numerical purposes, the limits of some of the functions at $M = 
0$ must be calculated separately, as explained in appendix \ref{AppB}. 

 \subsection{Still-expanding elliptic regions}

For $E < 0$, a similar result holds, but with one more refinement: 
depending on the value of $(t_2 - t_1)$, the final density will be either 
in the expansion phase or in the recollapse phase (and only in one of 
these phases). The dividing value of $(t_2 - t_1)$ will come out in the 
proof below. 

Let us assume that the $\eta$ of (\ref{EllEv}) is in $[0, \pi]$ for both 
values of $t_i$, so that the final density is still in the expansion phase 
of its evolution. (For $\eta \in [\pi, 2\pi]$, the solutions for $(t_i - 
t_B)$ are different, and they will be considered separately below.) The 
analogs of eqs. (\ref{tBiHyp}) and (\ref{t2t1Hyp}) are then: 

\begin{equation}
t_B = t_i - \frac {M} {(-2E)^{3/2}} \left[\arccos(1 + 2ER_i/M) -
\sqrt{1 - (1 + 2ER_i/M)^2}\right]
\end{equation}

 \noindent and

\begin{equation}   \label{psiEX0}
\psi_X(x) = 0,
\end{equation}

\noindent where this time

\begin{equation}   \label{psiEX}
x := - 2E/M^{2/3}, $$ $$\psi_X(x) := \arccos(1 - a_2x) - \sqrt{1 -
(1 - a_2x)^2} - \arccos(1 - a_1x) + \sqrt{1 - (1 - a_1x)^2} $$ $$-
(t_2 - t_1)x^{3/2} := \chi_X(x) - (t_2 - t_1)x^{3/2},
\end{equation}

\noindent the definitions of $a_i$ being still (\ref{psiH}).

The reasoning is entirely analogous to the one for (\ref{psiH}), but this
time the arguments of arccos must have absolute values not greater
than 1. This implies $x \leq 2/a_i$ for both $i$, and so, since $a_2
> a_1$

\begin{equation}   \label{xRangeEX}
0 \leq x \leq 2/a_2,
\end{equation}

\noindent which means: if there is any solution of (\ref{psiEX0}), then it 
will have the property (\ref{xRangeEX}). The two square roots in 
(\ref{psiEX}) will then also exist. Eq.~(\ref{xRangeEX}) is equivalent to 
the requirement that $(R,_t)^2$ (in (\ref{Rtsq}) with $\Lambda = 0$) is 
nonnegative at both $t_1$ and $t_2$. 

Note that

\begin{equation}   \label{psiEX_0_a2}
\psi_X(0) = 0, $$ 
$$\psi_X(2/a_2) = \pi - \arccos(1 - 2a_1/a_2) + 2\sqrt{a_1/a_2 - 
(a_1/a_2)^2} - (2/a_2)^{3/2} (t_2 - t_1),
\end{equation} 
\begin{equation}
\frac{d}{dx}\psi_X(x) = \psi_{X,x} = \sqrt{x} \left[\frac 
{{a_2}^{3/2}} {\sqrt{2 - a_2x}} -\frac {{a_1}^{3/2}} {\sqrt{2 - a_1x}} - 
\frac 32 (t_2 -t_1)\right].
\end{equation} 

\noindent In consequence of $a_2 > a_1$, the term in square
brackets is now an {\it increasing} function of $x$, and it
becomes $+\infty$ at $x = 2/a_2$ (which only means that $\psi_X(x)$
has a vertical tangent there). Hence, the term can go through zero
at most once, and it will do so when it is negative at $x = 0$,
i.e. when the opposite to (\ref{HypCondit}) holds

\begin{equation}   \label{EllCondit}
t_2 - t_1 > \frac {\sqrt{2}}3 \left({a_2}^{3/2} -
{a_1}^{3/2}\right).
\end{equation}

\noindent This means that the model must have expanded between $t_1$ and 
$t_2$ by less than the $E = 0$ model would have done. If (\ref{EllCondit}) 
does not hold, then $\psi_{X,x}$ is positive for all $x > 0$, which means 
that $\psi_X(x)$ is increasing and will not be zero for any $x > 0$. 
Hence, (\ref{EllCondit}) is a necessary condition for the existence of a 
solution of (\ref{psiEX0}). 

With (\ref{EllCondit}) fulfilled, $\psi_{X,x}(x)$ becomes negative for 
some $x > 0$, then goes through zero exactly once and then is positive all 
the way up to $x = 2/a_2$. This implies that $\psi_X(x)$ initially 
decreases below 0, then has exactly one minimum and is increasing up to 
the value (\ref{psiEX_0_a2}) at $x = 2/a_2$. Hence, $\psi_X(x)$ will have 
a zero for $x > 0$ only if $\psi_X(2/a_2) \geq 0$, i.e. if 

\begin{equation}   \label{StillXCondit}
t_2 - t_1 \leq (a_2/2)^{3/2} \left[\pi - \arccos(1 - 2a_1/a_2) +
2\sqrt{a_1/a_2 - (a_1/a_2)^2}\right].
\end{equation}

\noindent The inequality (\ref{StillXCondit}) is consistent with 
(\ref{EllCondit}), see appendix \ref{AppA}. Eqs. (\ref{EllCondit}) and 
(\ref{StillXCondit}) together are a necessary and sufficient condition for 
(\ref{psiEX0}) -- (\ref{psiEX}) to define a function $E(M) < 0$ for which 
$R(t_2, M)$ is still in the expansion phase of the model. 

 \subsection{Recollapsing elliptic regions}

The reasoning above applied only in the increasing branch of $R$
in (\ref{EllEv}). For the decreasing branch, where $\eta \in [\pi, 2\pi]$,
instead of (\ref{psiEX0}) -- (\ref{psiEX}) we obtain

 \begin{eqnarray}
   t_B & = & t_1 - \frac{M}{(-2E)^{3/2}} \left[ \arccos(1 + 2 E R_1 / M) -
\sqrt{1 - (1 + 2 E R_1 / M)^2}\; \right]   \nonumber \\
       & = & t_2 - \frac{M}{(-2E)^{3/2}} \left[ \pi + \arccos(- 1 - 2 E
R_2 / M) + \sqrt{1 - (1 + 2 E R_2 / M)^2}\; \right]
 \end{eqnarray}

\noindent and

\begin{equation}   \label{psiEC}
\psi_C = 0, \qquad {\rm where} $$ $$ \psi_C(x) := \pi + \arccos(- 1
+ a_2x) + \sqrt{1 - (1 - a_2x)^2} - \arccos(1 - a_1x) + \sqrt{1 -
(1 - a_1x)^2} $$ $$- (t_2 - t_1)x^{3/2}.
\end{equation}

\noindent The derivative of this is

\begin{equation}
\frac{d}{dx}\psi_C(x) = \psi_{C,x} = - \sqrt{x} \left[\frac 
{{a_2}^{3/2}} {\sqrt{2 -a_2x}} + \frac {{a_1}^{3/2}} {\sqrt{2 - a_1x}} + 
\frac 32 (t_2 -t_1)\right],
\end{equation} 

\noindent and is negative for all $x > 0$. Since

\begin{equation}
\psi_C(0) = 2\pi > 0,
\end{equation}

\noindent the solution of $\psi_C(x) = 0$ for $x > 0$ will
exist if and only if $\psi_C(2/a_2) \leq 0$, which translates into
the opposite of (\ref{StillXCondit}):

\begin{equation}   \label{RecollCondit}
t_2 - t_1 \geq (a_2/2)^{3/2} \left[\pi - \arccos(1 - 2a_1/a_2) +
2\sqrt{a_1/a_2 - (a_1/a_2)^2}\right].
\end{equation}

\noindent Thus the two densities can be connected by an $E < 0$ L-T 
evolution that is recollapsing at time $t_2$ if (\ref{RecollCondit}) is 
obeyed. 

Examples of functions $\psi_X(x)$ obeying or not obeying (\ref{EllCondit}) 
and (\ref{StillXCondit}), and of functions $\psi_C(x)$ obeying or not 
obeying (\ref{RecollCondit}) are shown in Figs. 2 and 3. 

 \begin{center}
 --------------------------- \\
 Figs. 2 and 3 go here \\
 ---------------------------
 \end{center}

 \subsection{Summary}

The above analysis considered only single world-lines, that is, single 
$M$ values.  We extend this to the whole of $\rho_i(M)$ by noting that 
$E(M)$ and $t_B(M)$ are arbitrary functions in the L-T model, and so 
continuous $\rho_i$ will generate continuous $E$ \& $t_B$.

The meaning of the limiting cases is now easy to understand. In 
(\ref{HypCondit}), at $M$ values where $t_2 - t_1 = 
(\sqrt{2}/3)({a_2}^{3/2} -{a_1}^{3/2})$, the final state results from the 
initial one by a parabolic ($E = 0$) evolution, so this $M$ value is on 
the boundary between an elliptic region and a hyperblic one.  
Eq.~(\ref{ParEv}) follows as the $E \rightarrow 0$ limit of (\ref{EllEv}) 
and of (\ref{HypEv}).  In (\ref{StillXCondit}) and (\ref{RecollCondit}), 
for $M$ values where the equality holds, the final state is exactly at the 
local moment of maximal expansion, separating a region of $\rho_2$ that is 
already recollapsing from one that is still expanding. 

However, when $E < 0$ it must be remembered that the signature of the 
metric requires that 

\begin{equation}   \label{Elimit}
E(M) \geq - 1/2,
\end{equation}

\noindent and so, once $E(M)$ has been calculated, (\ref{Elimit}) will 
have to be checked.  Note that the Friedmann model in standard coordinates 
has exactly this problem 
 --- with $2 E = -k r^2$ and $M = M_0 r^3$, blindly continuing through $r 
= 1$ will make $E < -1/2$ and $M > M_{Universe}$.  Indeed, given two 
uniform densities $\rho_2 < \rho_1$ that are appropriate for a closed 
Friedmann model, the integral (\ref{Rintegral}) can be extended to 
arbitrarily large $M$ \& $R$.  Thus the occurrence of $E = -1/2$ is not a 
problem, but rather an indication that the maximum of the spatial section 
has been reached.  One should record the values of $R_{max}$ and 
$M_{max}$, and then use 

\begin{equation}
  R^3_{{\rm max}i} - R^3(t_i, M)= \int_M^{M_{\rm max}} \frac{6}{\kappa 
\rho_i(M')} \, {\rm d}M'.
\end{equation} 

\noindent and distinguish the $M$ values beyond the maximum from those in 
front of it.

Another serious possibility is that shell crossings, where the 
density diverges and changes sign, may occur.  If they occur between $t_1$ 
\& $t_2$ the model evolution is unsatisfactory, but if they occur before 
$t_1$ or after $t_2$, this may not be of much concern.  The conditions on 
$E(M)$ \& $t_B(M)$ for avoiding them \cite{HeLa1985} must also be checked. 

All these considerations apply to the Friedmann limit, but it must
be remembered that in comparing models in a continuous Friedmann
family, one must not scale the curvature index $k$ to $+1$ or $-1$
when it is nonzero. The parameter $k$ is adapted to the initial
and final densities together with $M_0$. With $k$ scaled to $\pm
1$, taking the limit $k \to 0$ within the family becomes
impossible, and the inequalities (\ref{HypCondit}) and (\ref{EllCondit}) do
not come up.

In summary, for densities $\rho_2(M) < \rho_1(M)$ at times $t_2 >
t_1$ we have $a_2 > a_1$ where $a_i = R_i/M^{1/3}$, and writing
 \[
   \alpha = a_1/a_2
 \]
 the nature of the LT model that evolves between these states at a given
$M$ is: \\[5mm]
 ${}$~~~~~\fbox{ \parbox{145mm}{
 ${}$ \\
 \underbar{Hyperbolic $E > 0$}: \\
 If
 \[
   t_2 - t_1 < (\sqrt{2}\;a_2^{3/2}/3) (1 - \alpha^{3/2})
 \]
 then
 \[
   E = x M^{2/3}/2
 \]
 where $x$ solves
 \begin{eqnarray*}
   0 = \psi_H(x) & = & \sqrt{(1 + a_2 x)^2 - 1}\; - {\rm arcosh}(1 + a_2 x)
\\
   && - \sqrt{(1 + a_1 x)^2 - 1}\; + {\rm arcosh}(1 + a_1 x) - (t_2 -
t_1) x^{3/2}
 \end{eqnarray*}
 and
 \[
   t_B = t_i - \frac{1}{x^{3/2}} \left[ \sqrt{(1 + a_i x)^2 -1}\; - {\rm
arcosh}(1 + a_i x) \right]
 \]
 } }
 \\[3mm]
 ${}$~~~~~\fbox{ \parbox{145mm}{
 ${}$ \\
 \underbar{Parabolic $E = 0$}: \\
 If $(t_2 - t_1)$ is close to
 \[
   t_2 - t_1 = (\sqrt{2}\;a_2^{3/2}/3) (1 - \alpha^{3/2})
 \]
 then a series expansion gives
 \[
   E = x M^{2/3}/2
 \]
 where $x$ solves
 \begin{eqnarray*}
   0 = \psi_P(x) & \approx & \frac{\sqrt{2}\; x^{3/2}}{3} \left\{ a_2^{3/2} 
      \left( 1 - \frac{3}{20} a_2 x + \frac{9}{224} a_2^2 x^2 \right) 
      \right. \\ 
   && \left. - a_1^{3/2} 
      \left( 1 - \frac{3}{20} a_1 x + \frac{9}{224} a_1^2 x^2 \right) 
      - (t_2 - t_1) \right\}
 \end{eqnarray*}
 and
 \[
   t_B \approx t_i - \frac{\sqrt{2}\;}{3} a_i^{3/2} 
          \left( 1 - \frac{3}{20} a_i x + \frac{9}{224} a_i^2 x^2 \right) 
 \]
 } }
 \\[3mm]
 ${}$~~~~~\fbox{ \parbox{145mm}{
 ${}$ \\
 \underbar{Elliptic $E < 0$ and still expanding at $t_2$}: \\
 If
 \[
   (a_2/2)^{3/2} [\pi - \arccos(1 - 2 \alpha) + 2 \sqrt{\alpha -
\alpha^2}\;] > t_2 - t_1 > (\sqrt{2}\;a_2^{3/2}/3) (1 - \alpha^{3/2})
 \]
 then
 \[
   E = - x M^{2/3}/2
 \]
 where $x$ solves
 \begin{eqnarray*}
   0 = \psi_X(x) & = & \arccos(1 - a_2 x) - \sqrt{1 - (1 - a_2 x)^2}\; \\
   && - \arccos(1 - a_1 x) + \sqrt{1 - (1 - a_1 x)^2}\; - (t_2 - t_1)
x^{3/2}
 \end{eqnarray*}
 and
 \[
   t_B = t_i - \frac{1}{x^{3/2}} \left[ \arccos(1 - a_i x) - \sqrt{1 - (1
- a_i x)^2}\; \right]
 \]
 } }
 \\[3mm]
 ${}$~~~~~\fbox{ \parbox{145mm}{
 ${}$ \\
 \underbar{Elliptic $E < 0$ and at maximum expansion at $t_2$}: \\
 If $(t_2 - t_2)$ is close to
 \[
   t_2 - t_1 = (a_2/2)^{3/2} [\pi - \arccos(1 - 2 \alpha) + 2 \sqrt{\alpha
- \alpha^2}\;]
 \]
 then a series expansion gives
 \[
   E = - x M^{2/3}/2
 \]
 where $x$ solves
 \begin{eqnarray*}
   0 = \psi_M(x) & \approx & - 2^{3/2} (2 - a_2 x)^{1/2} 
      + \frac{2^{3/2}}{12} (2 - a_2 x)^{3/2} \\
   && + \pi - \arccos(1 - a_1 x) + \sqrt{a_1 x (2 - a_1 x)}\; 
      - x^{3/2} (t_2 - t_1) 
 \end{eqnarray*}
 and
 \begin{eqnarray*}
   t_B & = & t_1 - \frac{1}{x^{3/2}} \left[ \arccos(1 - a_1 x) - \sqrt{1 -
(1 - a_1 x)^2}\; \right] \\
   & \approx & t_2 - x^{-3/2} \left( \pi - 2^{3/2} (2 - a_2 x)^{1/2}
      + \frac{2^{3/2}}{12} (2 - a_2 x)^{3/2} \right) 
 \end{eqnarray*}
 } }
 \\[3mm]
 ${}$~~~~~\fbox{ \parbox{145mm}{
 ${}$ \\
 \underbar{Elliptic $E < 0$ and recollapsing at $t_2$}: \\
 If
 \[
   t_2 - t_1 > (a_2/2)^{3/2} [\pi - \arccos(1 - 2 \alpha) + 2 \sqrt{\alpha
- \alpha^2}\;]
 \]
 then
 \[
   E = - x M^{2/3}/2
 \]
 where $x$ solves
 \begin{eqnarray*}
   0 = \psi_C(x) & = & \pi - \arccos(- 1 + a_2 x) + \sqrt{1 - (1 - a_2
x)^2}\; \\
   && - \arccos(1 - a_1 x) + \sqrt{1 - (1 - a_1 x)^2}\; - (t_2 - t_1)
x^{3/2}
 \end{eqnarray*}
 and
 \begin{eqnarray*}
   t_B & = & t_1 - \frac{1}{x^{3/2}} \left[ \arccos(1 - a_1 x) - \sqrt{1 -
(1 - a_1 x)^2}\; \right] \\
       & = & t_2 - \frac{1}{x^{3/2}} \left[ \pi + \arccos(- 1 + a_2 x) +
\sqrt{1 - (1 - a_2 x)^2}\; \right]
 \end{eqnarray*}
 } }
 \\[3mm]
 It is easy to adapt the above for the case $\rho_2 > \rho_1$.  Clearly
any parabolic or hyperbolic regions would be collapsing%
 \footnote{
 For an expanding and a collapsing hyperbolic region to be contained in
the same smooth model, there would have to be an elliptic region between
them.  This is because the $\rho_2 < \rho_1$ region and the $\rho_2 >
\rho_1$ region must have a point between where $\rho_2 = \rho_1$.  The
only way this can be arranged without causing shell crossings is for the
elliptic region to be a
 Kruskal-like neck
 --- see \cite{Hell1987}
 }%
 .

 We conclude this section by stating the result as a theorem: 

 \noindent {\bf Theorem}~~~ Given any two times $t_1$ and $t_2 > t_1$, and 
any two spherically symmetric density profiles $0 < \rho_2(M) < \rho_1(M)$ 
defined over the same range of $M$, a L-T model can be found that evolves 
from $\rho_1$ to $\rho_2$ in time $t_2 - t_1$.  The inequalities 
(\ref{HypCondit})/(\ref{EllCondit}) and 
(\ref{StillXCondit})/(\ref{RecollCondit}) will tell which class of L-T 
evolution applies at each $M$ value.  The possibilities of shell crossings 
or excessively negative energies are not excluded, and must be separately 
checked for. 

 \section{Conditions for comoving extrema of density.}

 \setcounter{equation}{0}

Since we expect the central condensation to propagate outward into the
Friedmann background, we have to set up the initial conditions so that the
edge of the condensation is not comoving.  For this purpose, it is useful
to know the general conditions for comoving extrema of density.  We shall
now consider maxima and minima of $\rho$ in those domains where $\rho$ is
differentiable, and use $M$ as the radial coordinate.

From (\ref{rhoRM}) we see that extrema of $\rho$ will occur at those
values of $M$ where

\begin{equation}   \label{RcuMM0}
(R^3),_{MM} = 0.
\end{equation}

\noindent (This is a necessary condition only. Some of the
solutions of (\ref{RcuMM0}) will be inflection points of $\rho(t_1, M)$
rather than extrema, but the whole reasoning below will apply to
them,too.) For the case $E < 0$ we find from the second of (\ref{EllEv}):

\begin{equation}
(1 - \cos \eta) \eta,_M = \left[\frac {(- 2E)^{3/2}} {M}
\right],_M (t - t_B) - \frac {(- 2E)^{3/2}} {M} t_{B,M}.
\end{equation}

\noindent Using this in the first of (\ref{EllEv}) we find

\begin{equation}   \label{RcuMMeq}
(R^3),_{MM} = - \left(\frac{M^3} {8E^3}\right),_{MM}(1 - \cos
\eta)^3 $$ $$ - 6 \left(\frac{M^3} {8E^3}\right),_{M}(1 - \cos
\eta) \sin \eta \left\{\left[\frac {(- 2E)^{3/2}} {M} \right],_M
(t - t_B) - \frac {(- 2E)^{3/2}} {M} t_{B,M} \right\} $$ $$ - 3
\frac{M^3} {8E^3}\left(\frac {\sin^2\eta} {1 - \cos \eta} + \cos
\eta \right) \left\{\left[\frac {(- 2E)^{3/2}} {M} \right],_{M} (t
- t_B) - \frac {(- 2E)^{3/2}} {M} t_{B,M} \right\}^2 $$ $$  - 3
\frac{M^3} {8E^3}(1 - \cos \eta) \sin \eta \left\{\left[\frac {(-
2E)^{3/2}} {M} \right],_{MM} (t - t_B) \right. $$ $$ \left. - 2
\left[\frac {(- 2E)^{3/2}} {M}\right],_{M} t_{B,M} - \frac {(-
2E)^{3/2}} {M} t_{B,MM} \right\} = 0.
\end{equation}

\noindent This equation defines certain values of $M$, let us call
them $M = M_{\rm ex}$, at which $\rho$ may have extrema. We will
verify when they are comoving, i.e. when $M_{\rm ex}$ are
independent of time.

The Jacobian $\partial (M, \eta)/\partial (t, M)$ is nonzero
everywhere except those locations where

\begin{equation}
\eta,_t = 0 \qquad \Longrightarrow \qquad E = 0.
\end{equation}

\noindent Hence, everywhere else $M$ and $\eta$ can be considered
to be independent variables. Using the second of (\ref{EllEv}), we can
eliminate $(t - t_B)$ from (\ref{RcuMMeq}), and, since the $M_{\rm ex}$
obeying (\ref{RcuMMeq}) are assumed independent of $t$, what results is an
equation in $\eta$ with coefficients depending on $M$. The
coefficients of independent functions of $\eta$ all have to
vanish. This implies:

\begin{equation}   \label{CoMvMxCondit1}
\left[(- 2E)^{3/2}/M\right],_{M} = \left[(-
2E)^{3/2}/M\right],_{MM} = t_{B,M} = t_{B,MM} = 0,
\end{equation}

\begin{equation}   \label{CoMvMxCondit2}
\left(M^3/E^3\right),_{MM} = 0,
\end{equation}

\noindent all quantities being calculated at $M = M_{\rm ex}$. Eqs. 
(\ref{CoMvMxCondit1}) would have the same form in any coordinate system in 
which $M = M(r)$, but for (\ref{CoMvMxCondit2}), the coordinate $M$ is 
privileged; in other coordinates this equation would look less readable. 
Note that eqs. (\ref{CoMvMxCondit1}) -- (\ref{CoMvMxCondit2}) imply that 
at $M = M_{\rm ex}$ the functions $t_B$, $\left[(- 2E)^{3/2}/M\right]$ and 
$\left(M^3/E^3\right)$ agree with their Friedmann values (as determined by 
$E^{3/2}/M$ and $t_B$ at $M_{\rm ex}$) up to the second derivatives.  This 
means that the local density is at all times the same as that of the 
Friedmann model that matches on there. 

In brief, we have shown that if $M = M_{\rm ex}$ is an extremum of 
density, the density is differentiable at $M_{\rm ex}$, and the extremum 
is comoving, then (\ref{CoMvMxCondit1}) -- (\ref{CoMvMxCondit2}) are 
fulfilled. Conversely, if $\rho$ is differentiable at $M_{\rm ex}$, has an 
extremum there, and (\ref{CoMvMxCondit1}) -- (\ref{CoMvMxCondit2}) are 
fulfilled, then the extremum will be comoving. 

By the same method it may be verified that in the $E(M) > 0$ region, where 
(\ref{HypEv}) apply, the conditions for a comoving extremum are again of 
the form (\ref{CoMvMxCondit1}) -- (\ref{CoMvMxCondit2}), except that now 
the $(- 2E)$ in (\ref{CoMvMxCondit1}) is replaced by $(2E)$. 

For the parabolic case, (\ref{ParEv}), the condition $(R^3),_{MM} = 0$
reads

\begin{equation}
- 9 \left[ \left(2t_{B,M} + Mt_{B,MM}\right)\left(t - t_B\right) -
M{t_{B,M}}^2\right] = 0,
\end{equation}

\noindent and so the extremum will be comoving when $t_{B,M} =
t_{B,MM} = 0$ at $M = M_{\rm ex}$.

Note that for an arbitrary L-T perturbation inside an exactly Friedmannian 
exterior, the density at the boundary may be discontinuous, as in the 
Einstein-Straus \cite{EiSt1945, EiSt1946} configuration.  Furthermore, 
even if it is $C^1$ at an initial moment, it may develop a discontinuity. 
However, if the above conditions hold at the boundary, they ensure the 
density is $C^1$ through the boundary at all times. 

 \section{Numerical Example}

 \setcounter{equation}{0}

 \subsection{Scales in the background}

 The age of the universe is currently believed to be about $14$~Gyr.  If 
recombination temperature is $\sim 2700^\circ$~K, then $z = 2700/2.73 
\approx 1000$ then recombination happened at about $t_r = 3 \times 
10^5$~years when the density was about $10^{10}$ times the present 
density. 

However, in a $k = 0$ dust (Friedmann) model, $H_0 = 65$~km/s/Mpc implies 
$t_0 = 2/3H_0 = 10$~Gyr, which would put $t_r$ at $10^5$~yr. 

In an LT model that is close to parabolic even today, eqs
(\ref{EllEv}) \& (\ref{HypEv}) require $\sinh \eta - \eta \approx
\eta^3/6 \approx \eta - \sin \eta$, so we need $\eta^2 << 20$, say
$\eta < 0.4$.  For a Friedmann model
 \begin{equation}
   M_{F} = M_0 r^3 ~~,~~~~~~
   2 E_{F} = \pm 2 E_0 r^2 ~~,~~~~~~
   {t_B}_{F} = 0
 \end{equation}
 From (\ref{EllEv}) \& (\ref{HypEv}) again the limit on $E_0$ is
 \begin{equation}   \label{E0limit}
   \frac{(2 |E_0|)^{3/2} t_0}{M_0} \approx \frac{\eta^3}{6} <
      \frac{0.4^3}{6} \approx 0.01
 \end{equation}

 In a closed model, the maximum in the spatial sections
 --- the `equator' of the 3-sphere
 --- is at $r = r_m$ where $2 E = -1$, so
 \begin{equation}
   2 |E_0| r_m^2 = 1 ~~~~\Rightarrow~~~~ r_m = \frac{1}{\sqrt{2 |E_0|}\;}
\\
 \end{equation}
 at which point the areal radius today is
 \begin{equation}
   R_{m0} = \left( \frac{9 M_0}{2 \, (2 |E_0|)^{3/2}} \right)^{1/3} \,
t_0^{2/3}
 \end{equation}
 For hyperbolic models, there is no maximum radius, but $R_{m0}$ gives the
curvature scale.
 Another way to restrict $E_0$ is to specify that the horizon scale $c
t_0$ be much less than $R_{m0}$, say $t_0 < R_{m0}/8$, which gives a
restriction similar to (\ref{E0limit}).

 A third way to limit $E_0$ is to specify $\Omega_0 > 0.03$.  In a
hyperbolic model
 \begin{equation}
   \Omega = \frac{8 \pi \rho R^2}{3 \dot{R}^2} = \frac{2}{1 + \cosh(\eta)}
 \end{equation}
 Solving this for $\eta$ and using (\ref{HypEv}) again gives
 \begin{equation}
   \frac{(2 |E_0|)^{3/2} t_0}{M_0} < 0.016
 \end{equation}

 However, within a condensation, the evolution may be nowhere near
parabolic.

 \subsection{Past null cones, horizons and scales on the CMB sky}
 \label{HorizonScales}

 The physical radius of the past null cone in a $k = 0$ Friedmann dust model
with scale factor $S \propto t^{2/3}$ is
 \begin{equation}
   L(t) = S \int_t^{t_0} \, \frac{1}{S} \, dt
        = 3 c \, (t_0^{1/3} t^{2/3} - t)
 \end{equation}
 so an observed angular scale of $\theta$ on the CMB sky has a physical
size at recombination of
 \begin{equation}
   L_r = L(t) \, \theta = 3 c \, (t_0^{1/3} t_r^{2/3} - t_r) \, \theta
 \end{equation}
   The present day size of the observed structure
 --- assuming it doesn't collapse
 --- is merely scaled up by the ratio of scale factors
 \begin{equation}
   L_0 = L_r \frac{S_0}{S_r}
 \end{equation}
 The scales would be fairly similar in reasonable $k \neq 0$ models.

 To determine the condensed structures that correspond to a given present
day scale in the background, the mass $M_c$ of the condensation is
divided by the present day density $\rho_{b,0}$ of the Friedmann
background ($k = 0$ dust) and
 cube-rooted:
 \begin{equation}
   L_{c0} = \left( \frac{3 M_c}{4 \pi \rho_{b,0}} \right)^{1/3}
 \end{equation}
 Conversely, the mass associated with a given scale at a given time is
 \begin{equation}
   M_c = \frac{4 \pi L_c(t)^3 \rho_b(t)}{3}
 \end{equation}

 The particle (causal) horizon at any given time $t$ is
 \begin{equation}
   C = S \int_0^{t} \, \frac{1}{S} \, dt = 3 c t
 \end{equation}
 and the visual horizon is
 \begin{equation}
   V = S \int_{t_r}^{t} \, \frac{1}{S} \, dt = 3 c \, (t - t_r^{1/3}
t^{2/3})
 \end{equation}
 The particle horizon takes no account of inflation, and retains a dust
equation of state before recombination, so is only included as a rough
scale of interest.

 \subsection{Scales in the perturbation}

 We imagine that present day structures accreted their mass from a
background that was close to Friedmannian, and therefore the scale
of the matter that is destined to end up in a present day
condensation is fixed by its present day mass.

 The COBE data shows $\delta T/T \sim 10^{-5}$ on scales of $10^\circ$,
and the density perturbations are $\delta\rho/\rho = 3\delta T/T \leq 3
\times 10^{-5}$ \cite{Padm1996}. The power spectrum ${\cal P}(k) =
|\delta_{\bf k}|^2 = $ where $\delta \rho / \rho = \Sigma_{\bf k} \,
\delta_{\bf k} \, e^{i {\bf k . x}}$ is commonly approximated by ${\cal
P}(k) = A k e^{-k s}$, where the cut-off scale $s$ is small compared to
the Hubble scale.  This is just ${\cal P} = A k$ at longer wavelengths
(smaller $k$) \cite{Peeb1993}.  COBE's measurements had a resolution of
$\sim 10^\circ$, while BOOMERANG's and MAXIMA's were $\sim 0.2^\circ$.
These angular scales correspond to length scales of $2$~Mpc and $50$~kpc
at the time of decoupling, and thus to $2$~Gpc and $50$~Mpc today.  Thus
we are only just beginning to detect void scale perturbations in the CMB.
Although the magnitude of galaxy scale or even supercluster scale 
perturbations, are not yet directly constrained by observations, we will 
retain the figure of $\sim 10^{-5}$. 

 The scales associated with present day structures are summarised in the
following table.

 \begin{center}
 \begin{tabular}{l|c|c|c|c}
     & Radius      &             & Density       & Angle \\
     & today       & Mass        & of sphere     & on CMB\\
     & (kpc)       & ($M_\odot$) & ($\rho_b$)    & sky ($^\circ$) \\
 \hline
   star
     & $2 \times 10^{-11}$
                   & $1$         & $2 \times 10^{29}$
                                                 & $8 \times 10^{-7}$ \\
   globular cluster
     & $0.1$       & $10^5$      & $2 \times 10^5$
                                                 & $4 \times 10^{-5}$ \\
   galaxy
     & $15$        & $10^{11}$   & $6 \times 10^4$
                                                 & $4 \times 10^{-3}$ \\
   Virgo cluster
     & $2\,000$    & $2 \times 10^{13}$
                                 & $5$           &  $0.02$ \\
   Virgo supercluster
     & $15\,000$   & $5 \times 10^{14}$
                                 & $0.3$         &  $0.06$ \\
   Abell cluster (example)
     & $800$       & $10^{15}$   & $4\,000$      &  $0.08$ \\
 \hline
   void
     & $6 \times 10^4$
                   & ?           &               & 0.4 \\
 \hline
   recomb horizon
     & $280$       &             &               & 1.8 \\
   present horizon
     & $9.2 \times 10^6$
                   &             &               & 59 \\
   visual horizon
     & $8.9 \times 10^6$
                   &             &               & 57 \\
 \hline
   COBE resolution
     & $1.6 \times 10^6$
                   & $1.9 \times 10^{21}$
                                 & (1)           & 10 \\
   BOOM/MAX resolution
     & $3.1 \times 10^4$
                   & $1.5 \times 10^{16}$
                                 & (1)           & 0.2
 \end{tabular}
 \\[2mm]
 \parbox{145mm}{ \small
 {\bfseries Table 1.~~ Approximate scales associated with present day 
structures.}  Note that the horizons are those in $k = 0$ Friedmann models 
without inflation, as given in section \ref{HorizonScales}.  The masses 
associated with the resolution scales of COBE, MAXIMA \& BOOMERANG are 
obtained by assuming a density equal to the parabolic background value 
$\rho_b$, as indicated by `(1)' in the density coulmn.  Useful collections 
of data can be found at 
 {\tt http://www.obspm.fr/messier/},
 {\tt http://adc.gsfc.nasa.gov/adc/sciencedata.html}, and
 {\tt http://www.geocities.com/atlasoftheuniverse/supercls.html}.
 }
 \end{center}

 \subsection{Choice of units \& scales}

For the background Friedmann model, we choose the simplest case,
as its only purpose it to get the cosmic timescales \& densities
approximately right:
 \begin{equation}
   \Lambda = 0 ~~,~~~~~~~~ k = 0 ~~,~~~~~~~~ p = 0
 \end{equation}

 We will use geometric units such that $c = 1 = G$, and the remaining
scale freedom of GR is fixed by choosing units in which the present day
mass of the condensation being considered is 1.  The corresponding
geometric length and time units are then:
 \begin{equation}   \label{GeomUDef}
   M_G = 1 ~~~~\Rightarrow~~~~ L_G = M_G G / c^2 ~~,~~~~~~~~
      T_G = M_G G / c^3
 \end{equation}

 \subsection{The Model}

 The principal limitation of the L-T model in the post-recombination era
is the absence of rotation.  However, once rotation has become a
significant factor in the collapse process, there is already a well
defined structure.  Later on pressure and viscosity will become important.
Our interest is in generating highly condensed structures in a short
enough timescale, and these factors only come into play once collapse is
well underway.  Because of the lack of rotation etc, all of which tend to
delay or halt collapse, we expect our model to be rapidly collapsing
rather than stationary at the present day.

 We choose to model an Abell cluster:
 \begin{eqnarray}
   M_{Abell~Cluster} & = & 10^{15}~M_\odot \\
   R_{Abell~Cluster} & = & 800~\mbox{kpc}
 \end{eqnarray}

 From (\ref{GeomUDef}) and the above table the associated geometric units
are
 \begin{eqnarray}   \label{GeomUVal}
   1~M_G & = & M_{Abell~Cluster}   \nonumber \\
   1~L_G & = & 48~\mbox{pc}   \nonumber \\
   1~T_G & = & 156~\mbox{years}   \nonumber \\
   \rightarrow~~~~~~
   M_{Abell~Cluster} & = & 1~M_G   \nonumber \\
   R_{Abell~Cluster} & = & 16800~L_G   \nonumber \\
   t_2 & = & 6.4 \times 10^7~T_G
 \end{eqnarray}

 \subsubsection{Final profile}

 At $t_2 = 10$~Gyrs~$= 6.4 \times 10^7~T_G$, we specify the density
profile to be
 \begin{equation}
   \rho_2(M) = \rho_{b,2} \left( 7000 \, e^{- (4 M)^2} \right)
 \end{equation}
 which is shown in fig. 4.
 \begin{center}
 --------------------------- \\
 Fig. 4. goes here \\
 ---------------------------
 \end{center}
 Now the Friedmann density at $t_2$ is:
 \begin{equation}
   \rho_{b,2} = 1.3 \times 10^{-17}~M_G/L_G^{3} ~=~ 8 \times 10^{-
27}~\mbox{kg/m}^3
 \end{equation}
 so the radius in the Friedmann `background' that contains this mass is%
 \footnote{
 For backgrounds with $k \neq 0$, the radius that contains this mass would 
be adjusted slightly. 
 }%
 :
 \begin{equation}
   R_{F,2} = \left( \frac{3 M_{Abell~Cluster}}{4 \pi \rho_{b,2}}
\right)^{1/3} ~=~ 260\,000~L_G
 \end{equation}
 Thus we find
 \begin{equation}
   (R_2(M))^3 = \int_0^M \, \frac{3}{4 \pi \rho_2(M')} \, dM'
      = \frac{3}{224\,000 \sqrt{\pi}\; \rho_{b,2}} {\rm erfi}(4 M)
 \end{equation}
 as shown in fig. 5,
   \begin{center}
   --------------------------- \\
   Fig. 5 a, b. go here \\
   ---------------------------
   \end{center}
 and the resulting $\rho_2(R)$ is shown in fig. 6.
   \begin{center}
   --------------------------- \\
   Fig. 6. goes here \\
   ---------------------------
   \end{center}

 \subsubsection{Initial profile}

 At $t_1 = 100$~k\,years~$=~ 10^{-5}~t_2 ~=~ 641~T_G$ we specify the
density perturbation to have mass
 \begin{equation}
   M_1 = 10^{-2} M_{Abell~Cluster}
 \end{equation}
 and density enhancement
 \begin{equation}
   3 \times 10^{-5} \rho_{b,1}
 \end{equation}
 for which the chosen profile is:
 \begin{equation}
   \rho_1(M) = \rho_{b,1} \left( \frac{1.00003 \, (1 + 100 \, M)}{1 +
100.003 \, M} \right)
 \end{equation}
 as plotted in fig. 7.
   \begin{center}
   --------------------------- \\
   Fig. 7. goes here \\
   ---------------------------
   \end{center}
 The Friedmann density at $t_1$ is:
 \begin{equation}
   \rho_{b,1} = 1.3 \times 10^{-7}~M_G/L_G^{3} ~=~ 8 \times 10^{-
17}~\mbox{kg/m}^3
 \end{equation}
 and the radius in the `background' that contains the total mass is:
 \begin{equation}
   R_{F,1} = \left( \frac{3 M_{Abell~Cluster}}{4 \pi \rho_{b,1}}
\right)^{1/3} ~=~ 57\,000~L_G
 \end{equation}
 The resulting $R_1(M)$,
 \begin{equation}
   (R_1(M))^3 = \int_0^M \, \frac{3}{4 \pi \rho_1(M')} \, dM'
      = \frac{3}{4 \pi \rho_{b,1}} \left( M - \frac{0.00003}{100.003}
           \ln( 1 + 100 \, M) \right)
 \end{equation}
 is shown in fig. 8, and $\rho_1(R)$ in fig. 9.
   \begin{center}
   --------------------------- \\
   Figs. 8 \& 9. go here \\
   ---------------------------
   \end{center}

 \subsection{Model Results}

 A Maple program was written to generate the formulas and then
solve for $E(M)$ and $t_B(M)$ numerically, as explained in sec. 3.
The results are shown in figs. 10. \& 11.
   \begin{center}
   --------------------------- \\
   Figs. 10 \& 11. go here \\
   ---------------------------
   \end{center}
 We see that $E$ is of order $10^{-5}$ which gives a recollapse timescale 
of $10^7~T_G = 1.7 \times 10^9$~yr, so that the curvature in the 
condensation is of order $M t_2 / (2E)^{3/2} \sim 0.17$.  The bang time 
perturbation is of order $2~T_G = 300$~years, and is quite negligible. 

 Stricly speaking, an increasing $t_B$, $t_{B,M} > 0$, creates a shell 
crossing, but for such a slight variation in $t_B$, the shell crossing 
occurs very early on, long before $t_1$ when the model becomes valid. 

 The  `velocity' $R_{,t}$ 
 --- rate of change of the areal radius $R$ 
 --- would, in a homogeneous model, increase as $M^{1/3}$, so plotting 
$R_{,t}/M^{1/3}$, as in fig. 12, indicates the velocity perturbation, as a 
deviation from a constant value. 
   \begin{center}
   --------------------------- \\
   Fig. 12. goes here \\
   ---------------------------
   \end{center}
 In this case, the perturbation is within $3.10^{-5}$ for $0 < M < 0.6$, 
where $\rho_2$ is large, but increases to $8.10^{-4}$ in the near vacuum 
region $0.6 < M < 1$.  This slight excess is most likely due to choosing a 
$\rho_2(M)$ that falls off too fast outside the condensation, requiring a 
too strongly hyperbolic evolution that expands too rapidly.  It is likely 
that this same effect requires the slightly increasing $t_B$ to keep 
$\rho_1(M)$ almost flat in these outer regions. 

 As a
 cross-check, these derived functions were used in a separate MATLAB 
program that plots the evolution of a 
 L-T model, given its arbitrary functions.  (The appropriate form of the 
evolution equations is given in appendix \ref{AppB}.)  The initial and 
final density profiles were recovered to high accuracy.  The resulting 
density evolution is shown in fig. 13. 
   \begin{center}
   --------------------------- \\
   Fig. 13. goes here \\
   ---------------------------
   \end{center}

 \section{Conclusions}

 \setcounter{equation}{0}

 \begin{itemize}

 \item   We proved that an L-T model can be found to evolve any initial
density profile on a constant time slice, to any final density profile a
given time later.  Although it can't be guaranteed the resulting model is
free of physical singularities such as shell crossings, and the occurrence 
of a large negative energy ($E = -1/2$) must be handled correctly by 
changing to $M$ \& $R$ decreasing, our numerical experiments indicate that 
realistic choices of the two density profiles and the time difference are 
likely to generate reasonable models. 

 \item   Our numerical example created an Abell cluster in a realistic
timescale.  It started from recombination, with a density perturbation 
involving a small amount of mass,  having $\delta \rho / \rho \sim 3.10^{-
5}$.  It then `accreted' most of its final mass.  In fact this `accretion' 
consists of lower expansion rates near the centre, and more rapid 
expansion rates further out.  Only at late stages does actual collapse 
begin at the centre.  The initial velocity perturbation cannot be chosen 
if the initial and final densities are chosen.  It turned out to be 
$\delta v / v \sim 3.10^{-5}$ within the future condensation and $\sim 
8.10^{-4}$ in the future vacuum region.  The relativiely large value in 
the outer regions is probably due to choosing the final density profile, 
$\rho_2$ to fall off more rapidly than ideal. 

 \item   The preceding two points 
 --- the theorem plus the numerical example 
 --- demonstrate that the L-T model provides a very reasonable description 
of post-recombination structure formation. 

 \item   These two points also indicate that post recombination structure 
formation in a dust universe has an important kinematical component 
 --- the initial distribution of velocities has as much bearing on whether 
or not a condensation forms and gravity magnifies density fluctuations, as 
the initial density distribution.  These initial distributions of density 
and velocity are generated by the functions $E(M)$ and $t_B(M)$, i.e. 
coded in the initial conditions. 

 \item   Further numerical examples for structure formation on a variety
of scales within the L-T model are under investigation.

 \item   We also obtained the conditions for a density maximum or minimum 
or shoulder to be comoving.  Since these are resitrictions on the L-T 
arbitrary functions, it is evident that extrema are in general moving 
through the fluid, as argued in \cite{ElHMa1990}, and are not comoving.  
In other words present day density maxima are not likely to be on the same 
worldlines as initial density maxima.

 \end{itemize}

 \noindent {\bf Acknowledgements}~~~

A. K. is grateful to the Department of Mathematics and Applied
Mathematics, University of Cape Town, for hospitality and perfect
working conditions; and to Charles Hellaby for being a most caring
host. We are grateful to G. F. R. Ellis for useful comments on the
first draft.

 \setcounter{secnumdepth}{0}

 \setcounter{secnumdepth}{3}
 \appendix

 \section{Consistency of the inequalities (\ref{EllCondit}) and 
(\ref{StillXCondit}).} 

 \label{AppA}

 \setcounter{equation}{0}

The inequalities (\ref{EllCondit}) and (\ref{StillXCondit}) will be 
consistent if the right-hand side of (\ref{EllCondit}) is smaller than the 
r.h.s. of (\ref{StillXCondit}) in the whole range of $a_1$ and $a_2$, i.e. 
when 

\begin{equation}   \label{CompInequal}
\frac {\sqrt{2}}3 \left({a_2}^{3/2} - {a_1}^{3/2}\right) <
(a_2/2)^{3/2} \left[\pi - \arccos(1 - 2a_1/a_2) + 2\sqrt{a_1/a_2 -
(a_1/a_2)^2}\right]
\end{equation}

\noindent Defining $y = a_1/a_2$, and recalling that $a_2 > a_1 >
0$ by assumption, the above is equivalent to

\begin{equation}
f(y) > 0 \qquad {\rm for\ all\ } 0 < y < 1, \  {\rm where} $$ $$
f(y) := \pi - \arccos(1 - 2y) + 2\sqrt {y - y^2} - \frac 43(1 -
y^2).
\end{equation}

\noindent Now observe that

\begin{equation}
f(0) = \pi - \frac 43 > 0, \qquad f(1) = 0, $$ $$ \frac {{\rm d}f}
{{\rm d}y} = 2\frac{\sqrt{y}}{\sqrt{1 - y^2}} (\sqrt{1 - y} - 1) <
0.
\end{equation}

\noindent Hence, $f(y)$ is monotonically decreasing from $f(0) >
0$ to 0 and so is positive for all $0 < y < 1$, which proves
(\ref{CompInequal}).

 \section{Numerical considerations}

 \label{AppB}

 \setcounter{equation}{0}

 \subsection{Limiting values at $M = 0$.}

 Several of the quantities considered in this paper have the value 0 at 
the centre of symmetry, where $M = 0$.  The variables used in the proof of 
the theorem, $a_i$ and $x$, have finite limits as $M \to 0$.  For 
numerical programs, these limiting values have to be provided explicitly. 

 The values of $a_i$ at $M = 0$ follow very easily.  Since $a_i = R(t_i, 
M)/M^{1/3}$ and $R^3 = \int_0^M \frac {6}{\kappa \rho(t, x)} {\rm d} x$, 
we have, applying the de l'Hospital rule in the third step 

\begin{equation}
a_i(0) = \lim_{M \to 0} \frac {R(t_i, M)}{M^{1/3}} = \left[\lim_{M
\to 0} \left( \frac {R^3}{M} \right)\right]^{1/3} $$ $$=
\left(\lim_{M \to 0} \frac {\rm d}{{\rm d}M} R^3 \right)^{1/3} =
\left(\lim_{M \to 0} \frac {6} {\kappa \rho} \right)^{1/3} =
\left(\frac {6}{\kappa \rho(t_i, 0)} \right)^{1/3}.
\end{equation}

 The variable $x$ comes out nonzero automatically when nonzero values of 
$a_i(0)$ are used in the program; this follows from the proof of the 
theorem in sec. \ref{TheoremSection}.

 \subsection{Practical variables}

 In practice, it was convenient to define $\alpha = a_1/a_2$, $z_i = 
t_i/a_2$, and $y = a_2 x$, and then solve the various $\psi = 0$ equations 
in terms of the variable $y$, since $a_2$ was in general quite large, 
whereas $0 < a x < 2$ in elliptic regions, and $a x$ was at most $200$ in 
a quite strongly hyperbolic region. 

 \subsection{Reconstructing the model evolution}

 For reconstructing the evolution of the model, it is convenient to 
 re-write the 
 L-T solutions in terms of $x$ \& $a$: \\
Elliptic:
 \[
   R = M^{1/3} \frac{(1 - \cos \eta)}{x}
   ~,~~~~~~
   t - t_B = \frac{(\eta - \sin \eta)}{x^{3/2}}
   ~,~~~~~~
   \dot{R} = M^{1/3} \sqrt{x \left( \frac{2}{1 - \cos \eta} - 1 \right)}\; 
 \]
Parabolic or close to it:
 \begin{eqnarray*}
   R & = & M^{1/3} \left( \frac{9}{2} \right)^{1/3} (t - t_B)^{2/3} 
      \left( 1 + \frac{x}{20} [6 (t - t_B)]^{2/3} \right.   \nonumber \\
   && \left. - \frac{3 x^2}{2800} [6 (t - t_B)]^{4/3} 
             + \frac{23 x^3}{504000} [6 (t - t_B)]^2 \right)
 \end{eqnarray*}
 \begin{equation}
   \dot{R} = M^{1/3} \sqrt{ 4 [6 (t - t_B)]^{-2/3}
      \left( 1 + \frac{x}{5} [6 (t - t_B)]^{2/3}
             + \frac{x^2}{280} [6 (t - t_B)]^{4/3}
             - \frac{x^3}{3600} [6 (t - t_B)]^2 \right) }\;
 \end{equation} 
Hyperbolic:
 \[
   R = M^{1/3} \frac{(\cosh \eta - 1)}{x}
   ~,~~~~~~
   t - t_B = \frac{(\sinh \eta - \eta)}{x^{3/2}}
   ~,~~~~~~
   \dot{R} = M^{1/3} \sqrt{x \left( \frac{2}{\cosh \eta - 1} + 1 \right)}\; 
 \]
 In all cases the density is
 \[
   \rho = \frac{1}{4 \pi a^2 \left( a/3 + M a_{,M} \right)}
 \]

 \

 \newpage

 \centerline {\bf FIGURES}

 {\small

 ${}$ \hspace{15mm}
 \includegraphics[scale = 0.75]{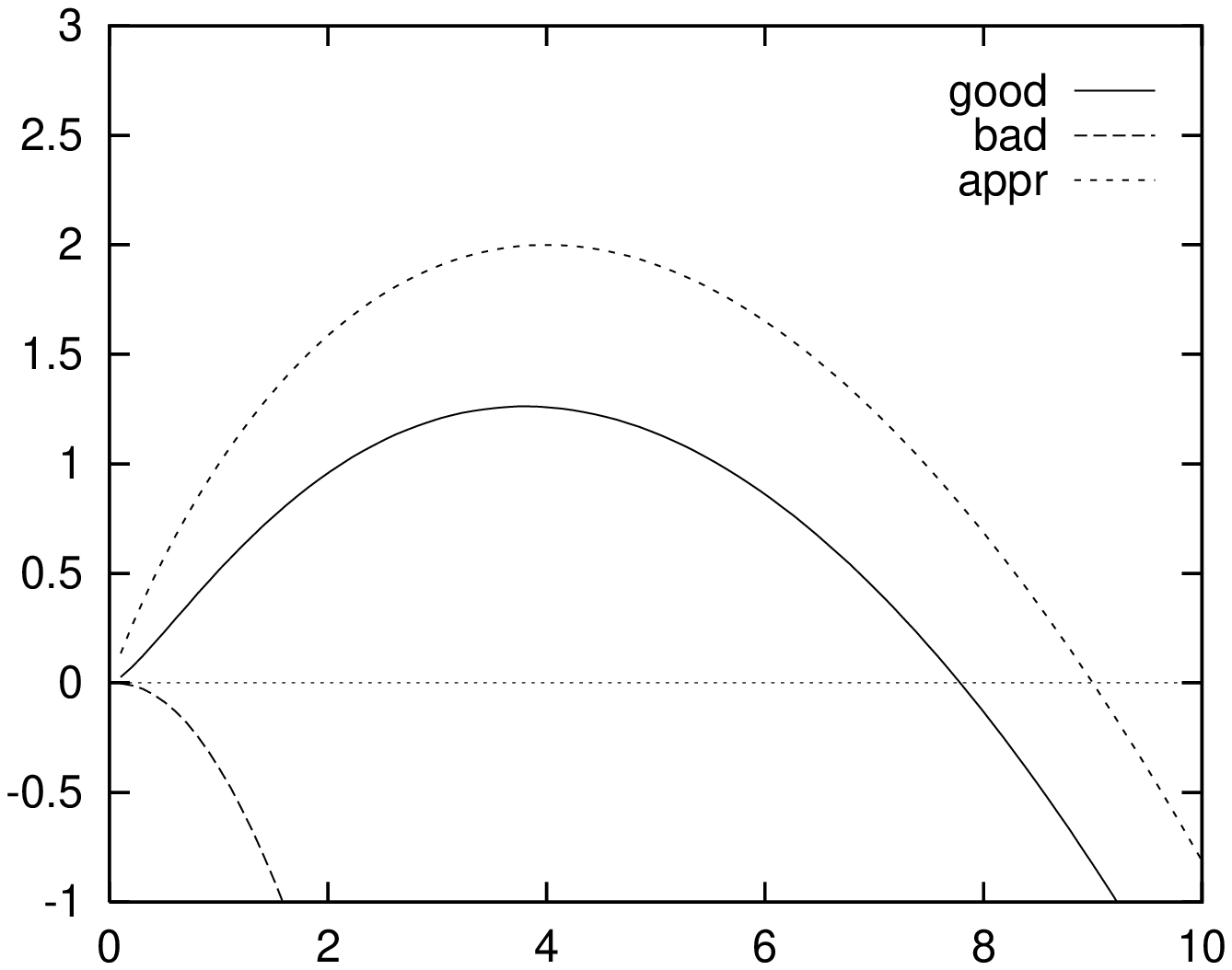} \\
Fig. 1. The function $\psi_H(x)$ for the case $E > 0$ with 
(\ref{HypCondit}) fulfilled (middle curve) and with (\ref{HypCondit}) not 
fulfilled (lower curve). The upper curve is the approximating function 
$\psi_{A}(x)$ from (\ref{psiA}). The parameters $(a_1, a_2, t_2 - t_1)$ 
are $(1, 2.5, 0.5)$ for the two upper curves and $(1, 2.5, 1.4)$ for the 
lower one. 

 ${}$ \hspace{15mm}
 \includegraphics[scale = 0.75]{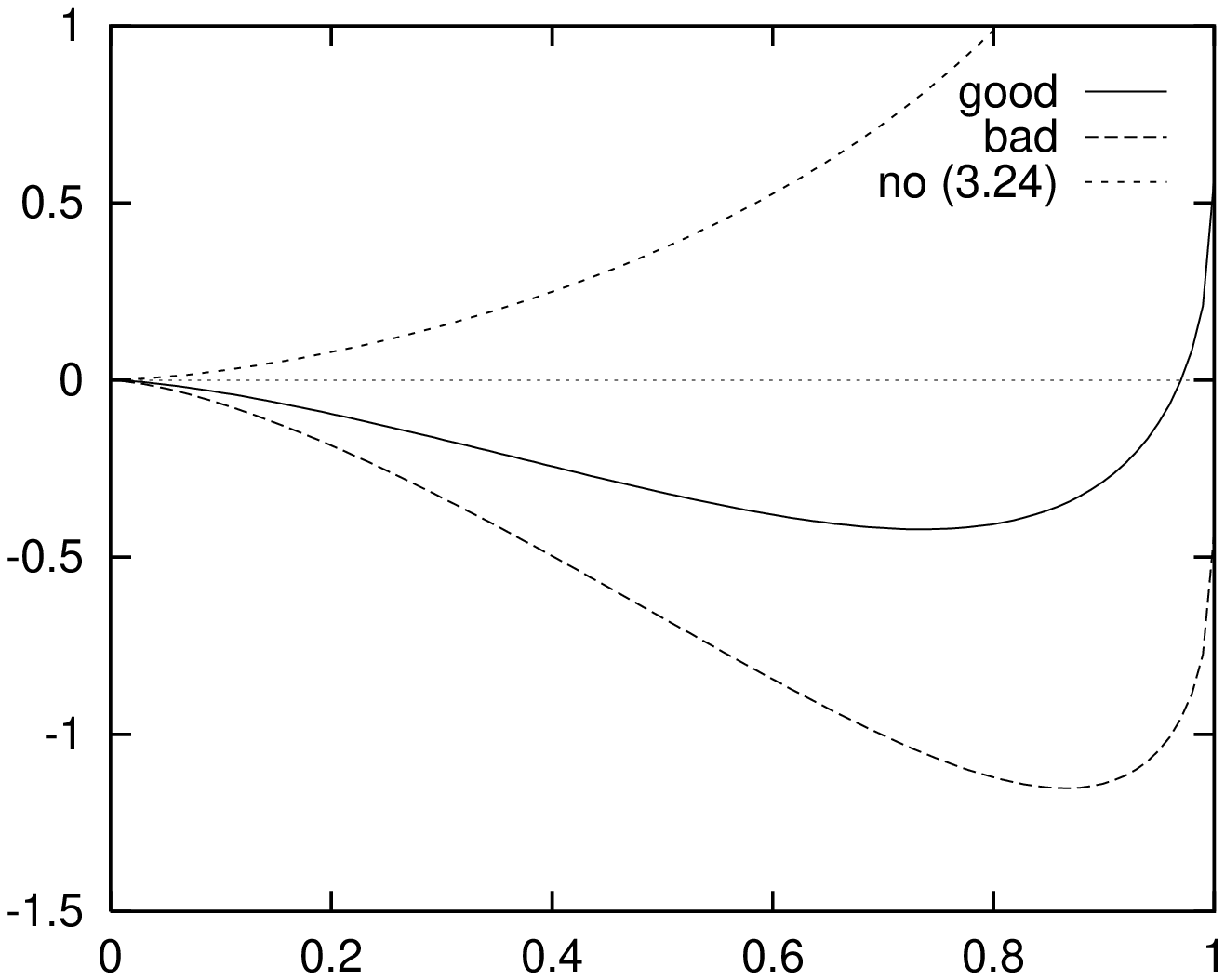} \\
Fig. 2. The function $\psi_X(x)$ for the case $E < 0$ with 
(\ref{EllCondit}) and (\ref{StillXCondit}) fulfilled (middle curve), with 
(\ref{EllCondit}) fulfilled and (\ref{StillXCondit}) not fulfilled (lower 
curve), and with (\ref{EllCondit}) not fulfilled (upper curve). The 
parameters $(a_1, a_2, t_2 - t_1)$ are $(1, 2, 2)$ for the middle curve, 
$(1, 2, 3)$ for the lower curve, and $(1, 2, 0.05)$ for the upper curve. 

 \newpage
 ${}$ \hspace{15mm}
 \includegraphics[scale = 0.75]{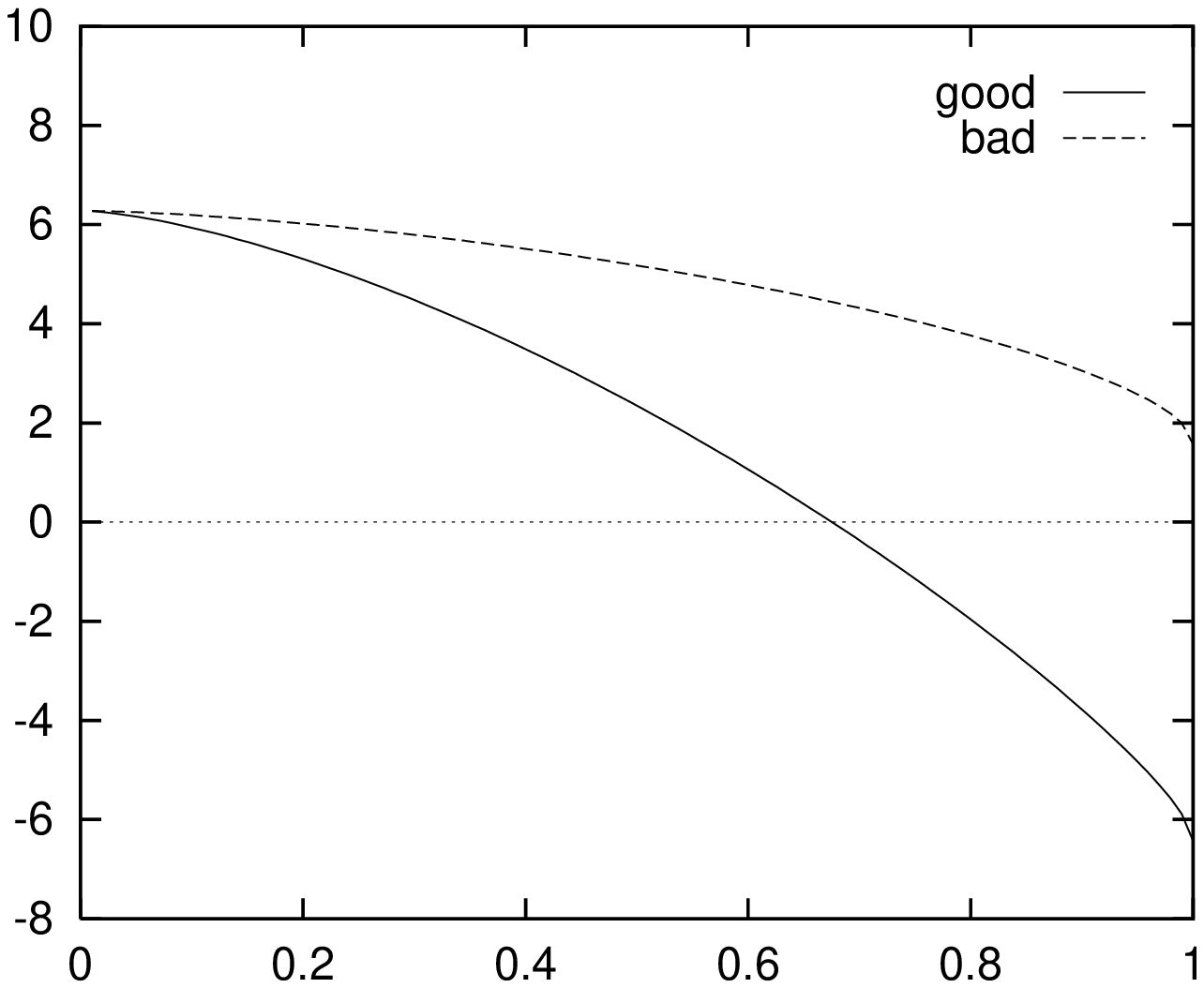} \\
Fig. 3. The function $\psi_C(x)$ from (\ref{psiEC}) obeying 
(\ref{RecollCondit}) (solid curve) and not obeying it (dashed curve). The 
parameters $(a_1, a_2, t_2 - t_1)$ are $(1, 2, 9)$ for the first curve and 
$(1, 2, 0.2)$ for the other one. 

 ${}$ \hspace{15mm}
 \includegraphics[angle=270, scale = 0.56]{rho2M_.eps} \\
 Fig.4~~~The chosen density profile $\rho_2(M)/\rho_{b,2}$ as a multiple
of the `background' density, for an Abell cluster at time $t_2 = 10$~Gyr.
The axes are in geometric units such that $M_{Abell~cluster}=1$, as given
in (\ref{GeomUDef}) \& (\ref{GeomUVal}).

 \newpage
 ${}$ \hspace{15mm}
 \includegraphics[angle=270, scale = 0.56]{R2M_.eps}
   \vspace*{-88mm} \\
   ${}$ \hspace{15mm}
   ${}$\hspace*{25mm}
 \includegraphics[angle=270, scale = 0.375]{R2Mb_.eps}
   \vspace*{30mm} \\
 Fig. 5~~~The areal radius $R_2(M)$ at time $t_2$, that results from the
chosen $\rho_2(M)$.  The inset shows an enlargement of the curve near
small $M$.  The axes are in geometric units.

 ${}$ \hspace{15mm}
 \includegraphics[angle=270, scale = 0.56]{lnrho2R_.eps} \\
 Fig. 6~~~The density profile $\rho_2/\rho_{b,2}$ against areal radius
$R_2$.  The axes are in geometric units.

 \newpage
 ${}$ \hspace{15mm}
 \includegraphics[angle=270, scale = 0.56]{rho1M_.eps} \\
 Fig. 7~~~The density profile $\rho_1(M)/\rho_{b,1}$ chosen for the
initial perturbation at time $t_1$.  The axes are in the geometric units
of (\ref{GeomUDef}) \& (\ref{GeomUVal}).

 ${}$ \hspace{15mm}
 \includegraphics[angle=270, scale = 0.56]{R1M_.eps} \\
 Fig. 8~~~Areal radius $R_1(M)$ at time $t_1$, that is obtained from
$\rho_1(M)$.  The axes are in geometric units.

 \newpage
 ${}$ \hspace{15mm}
 \includegraphics[angle=270, scale = 0.56]{rho1R_.eps} \\
 Fig. 9~~~Density profile $\rho_1/\rho_{b,1}$ against areal radius $R_1$.
The axes are in geometric units.

 ${}$ \hspace{15mm}
 \includegraphics[scale = 0.64]{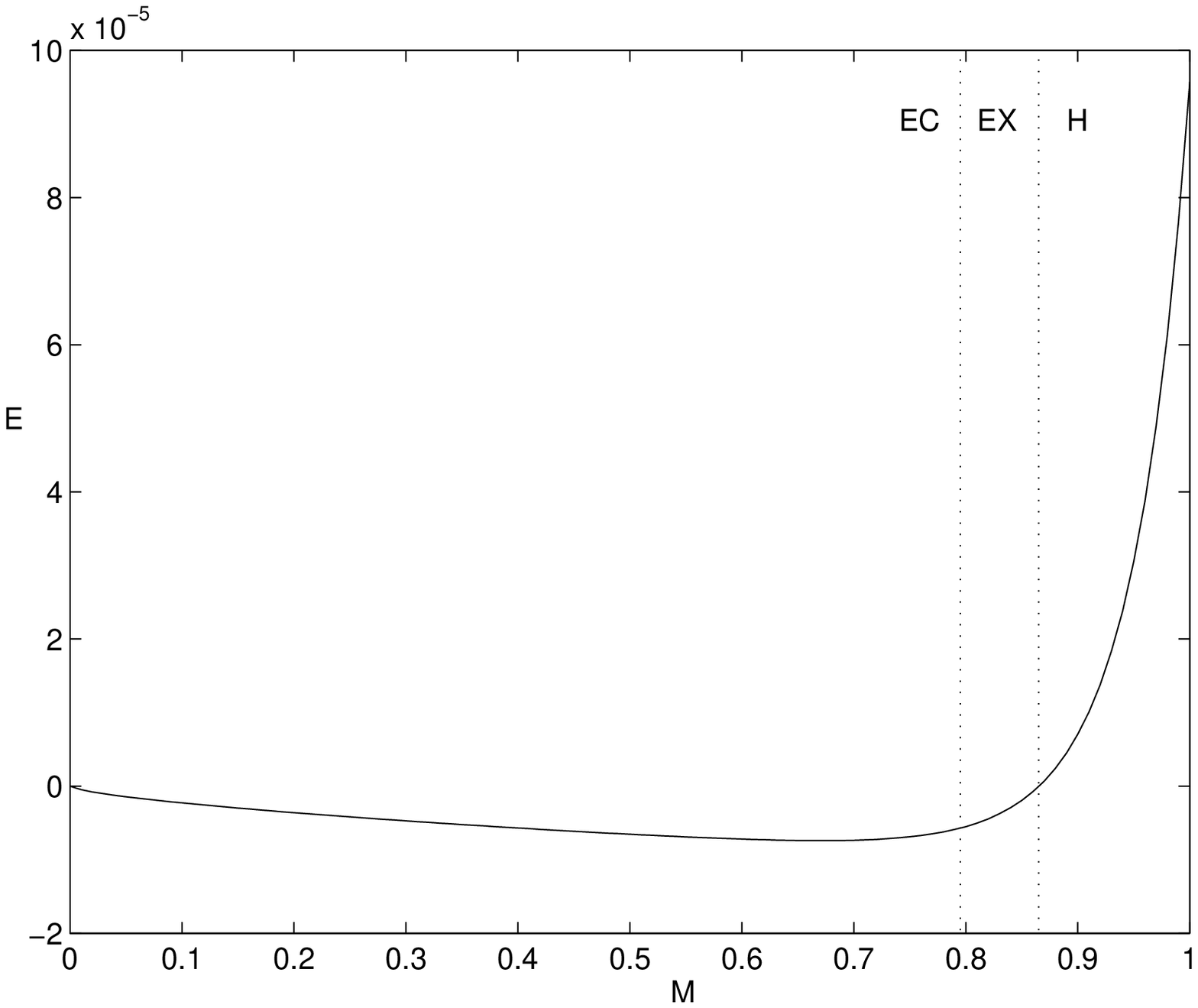} \\
 Fig. 10~~~The L-T energy function $E(M)$ obtained from solving for the
 L-T model that evolves between $\rho_1(M)$ and $\rho_2(M)$.  The axes are
in geometric units.  The symbols ``EC", ``EX" \& ``H" indicate regions 
that are respectively elliptic and recollapsing at $t_2$ (EC), elliptic 
and still expanding at $t_2$ (EX), and hyperbolic (H). 

 \newpage
 ${}$ \hspace{15mm}
 \includegraphics[scale = 0.64]{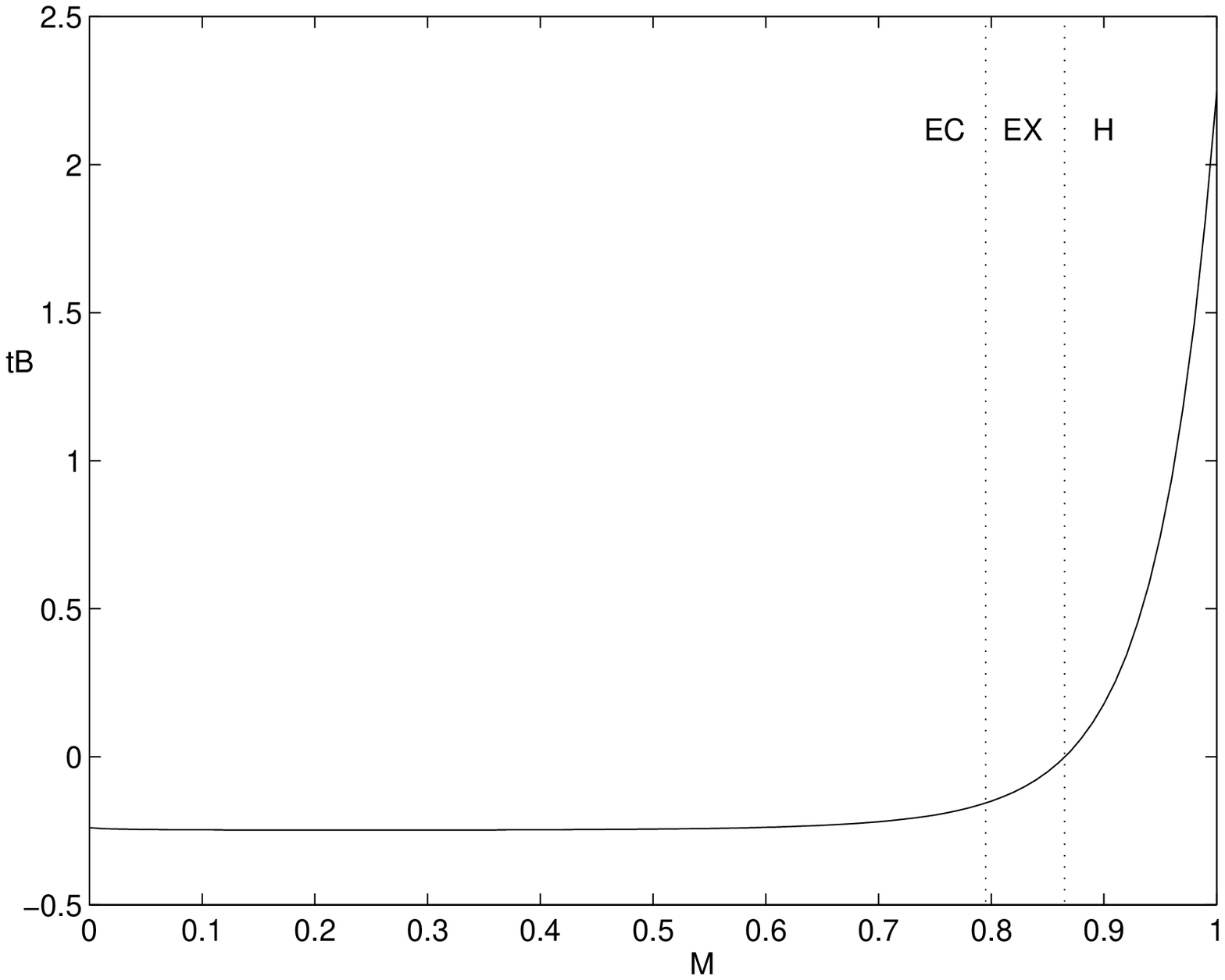} \\
 Fig. 11~~~The L-T bang time function $t_B(M)$ obtained from solving for
the L-T model that evolves between $\rho_1(M)$ and $\rho_2(M)$.  The axes
are in geometric units.

 ${}$ \hspace{15mm}
 \includegraphics[scale = 0.64]{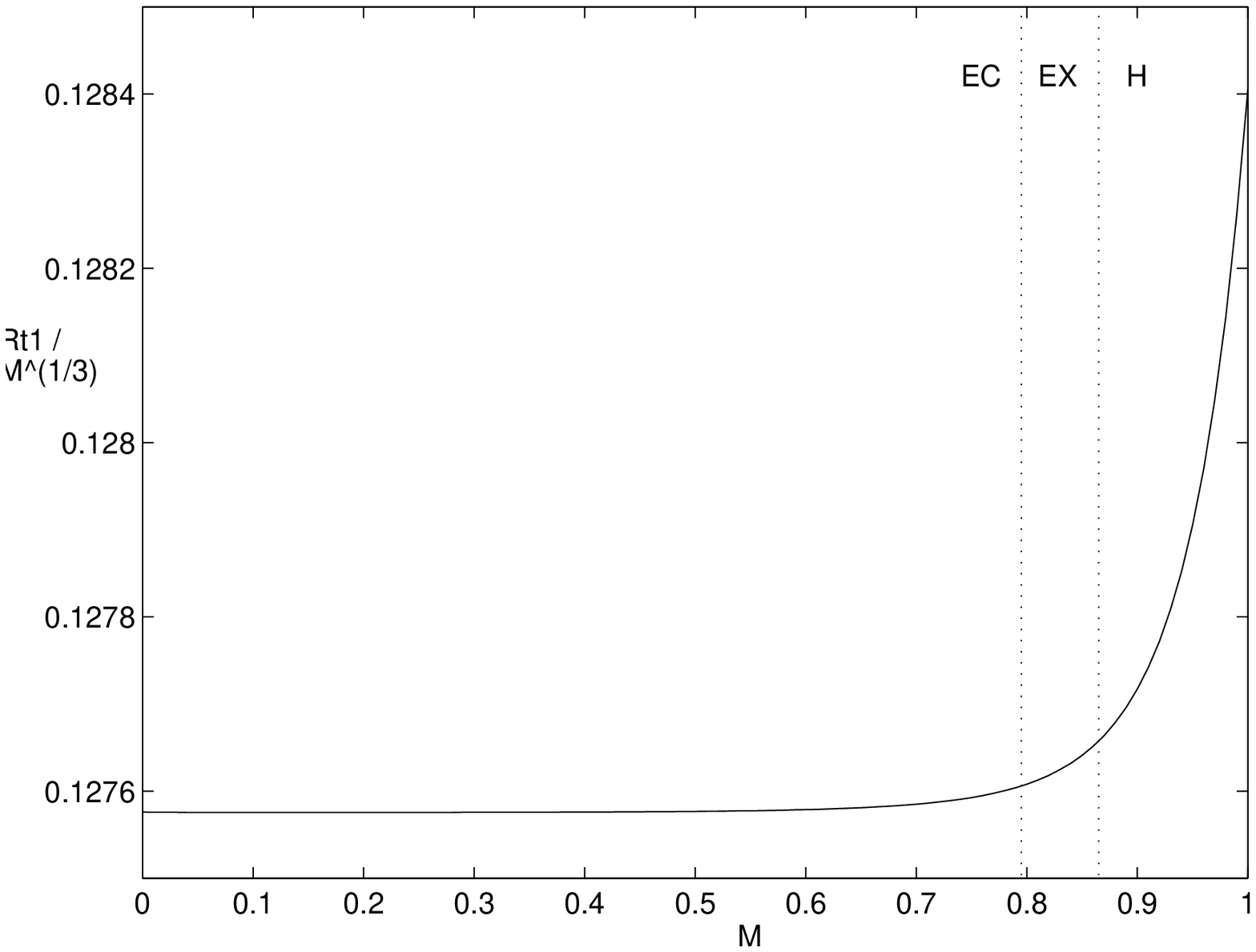} \\
 Fig. 12~~~The velocity perturbation $\dot{R}/M^{1/3}$ at time $t_1$.  A 
constant value would indicate no perturbation.  The axes are in geometric 
units.  

 \newpage
 ${}$ \hspace{15mm}
 \includegraphics[scale = 0.64]{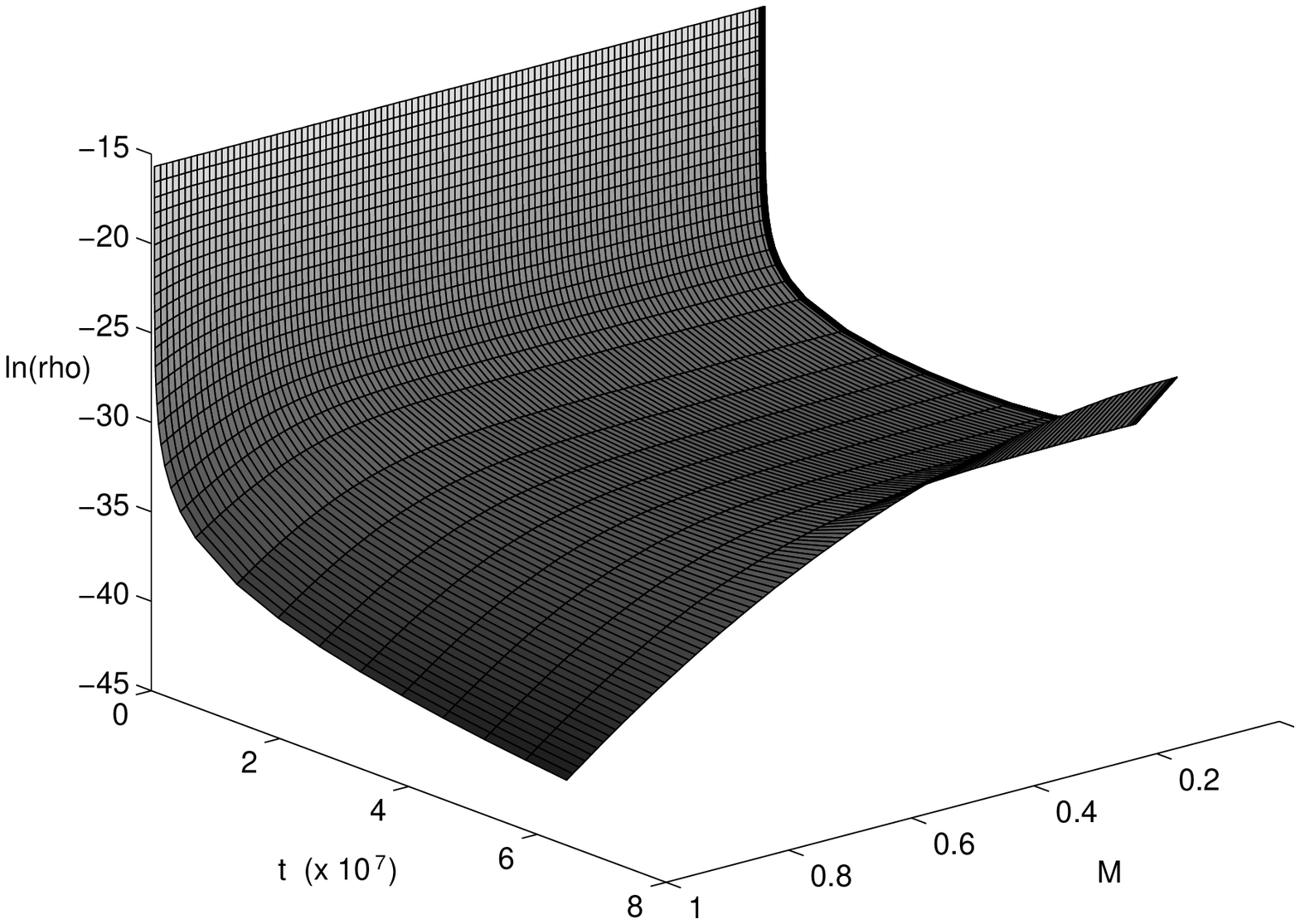} \\
 Fig. 13~~~The evolution of $\rho(t, M)$ for the derived L-T model.  The 
axes are in the geometric units of (\ref{GeomUDef}) \& (\ref{GeomUVal}).  
In the range $0 < M < 0.795$ the evolution is elliptic and already 
recollapsing at time $t_2$, in $0.795 < M < 0.865$ it is elliptic but 
still expanding at $t_2$, and for $M > 0.865$ it is hyperbolic.  In 
practice, recollapse would be halted at some point by the effects of 
pressure, rotation, etc.  The initial and final density profiles 
calculated at times $t_1$ and $t_2$ coincide with those originally chosen 
and shown in figures 7 and 4 respectively. 

 }

\end{document}